\documentclass[aps,pra,superscriptaddress,twocolumn,showpacs,amsmath,floatfix]{revtex4-2}

\usepackage{float}
\usepackage{amsmath}
\usepackage{graphicx}
\usepackage{color}
\usepackage{textcomp}
\usepackage{adjustbox}
\usepackage[colorlinks,linkcolor=blue]{hyperref}
\usepackage{tikz}

\tikzset{every picture/.style={line width=0.75pt}} 



\usepackage{babel}

\begin{document}
\title{Entanglement entropy scaling of noisy random quantum circuits in two dimensions}	
\author{Meng Zhang}
\affiliation{CAS Key Laboratory of Quantum Information, University of Science and Technology of China, Hefei 230026, Anhui, China}
\affiliation{Synergetic Innovation Center of Quantum Information and Quantum Physics, University of Science and Technology of China, Hefei, 230026, China}
\affiliation{Hefei National Laboratory, Hefei 230088, China}
\author{Chao Wang}
\affiliation{CAS Key Laboratory of Quantum Information, University of Science and Technology of China, Hefei 230026, Anhui, China}
\affiliation{Synergetic Innovation Center of Quantum Information and Quantum Physics, University of Science and Technology of China, Hefei, 230026, China}
\affiliation{Hefei National Laboratory, Hefei 230088, China}
\author{Shaojun Dong}
\affiliation{CAS Key Laboratory of Quantum Information, University of Science and Technology of China, Hefei 230026, Anhui, China}
\affiliation{Synergetic Innovation Center of Quantum Information and Quantum Physics, University of Science and Technology of China, Hefei, 230026, China}
\affiliation{Institute of Artificial Intelligence, Hefei Comprehensive National Science Center}
\author{Hao Zhang}
\affiliation{CAS Key Laboratory of Quantum Information, University of Science and Technology of China, Hefei 230026, Anhui, China}
\affiliation{Synergetic Innovation Center of Quantum Information and Quantum Physics, University of Science and Technology of China, Hefei, 230026, China}
\affiliation{Hefei National Laboratory, Hefei 230088, China}
\author{Yongjian Han}%
\email{smhan@ustc.edu.cn}
\affiliation{CAS Key Laboratory of Quantum Information, University of Science and Technology of China, Hefei 230026, Anhui, China}
\affiliation{Synergetic Innovation Center of Quantum Information and Quantum Physics, University of Science and Technology of China, Hefei, 230026, China}
\affiliation{Hefei National Laboratory, Hefei 230088, China}
\affiliation{Institute of Artificial Intelligence, Hefei Comprehensive National Science Center}
\author{Lixin He}%
\email{helx@ustc.edu.cn}
\affiliation{CAS Key Laboratory of Quantum Information, University of Science and Technology of China, Hefei 230026, Anhui, China}
\affiliation{Synergetic Innovation Center of Quantum Information and Quantum Physics, University of Science and Technology of China, Hefei, 230026, China}
\affiliation{Hefei National Laboratory, Hefei 230088, China}
\begin{abstract}

Whether noisy quantum devices without error correction can provide quantum advantage over classical computers is a critical issue of current quantum computation. In this work, the random quantum circuits, which are used as the paradigm model to demonstrate quantum advantage, are simulated with depolarizing noise on experiment relevant two-dimensional architecture. With comprehensive numerical simulation and theoretical analysis, we find that the maximum achievable operator entanglement entropy, which indicates maximal simulation cost, has area law scaling with the system size for constant noise rate. On the other hand, we also find that the maximum achievable
operator entanglement entropy has power law scaling with the noise rate for fixed system size, and the volume law scaling can be obtained
only if the noise rate decreases when system size increase.
\end{abstract}

\maketitle

\section{Introduction}
Quantum computer is expected to have exponential speedup in certain important problems, such as integer factorization \cite{shoralgorithms} and quantum dynamics \cite{RevModPhys.86.153}, against classical computers.
However, implementing aforementioned algorithms require large-scale full fault-tolerant quantum computers, which are beyond the capabilities of the existing technology. To demonstrate the quantum advantage in Noisy Intermediate-Scale Quantum (NISQ) \cite{Preskill2018quantumcomputingin} devices, which are available currently, is an important issue for current quantum computation.

The random circuit sampling \cite{boixo2018characterizing} and boson sampling \cite{aaronson2011computational} problems are proposed and theoretically proved (with some reasonable assumption) as the prominent computing problems to achieve the quantum advantage. More importantly, several experiment groups have implemented the random circuit sampling and claimed quantum supreme. Specifically, the first quantum supremacy in random quantum circuits (RQCs) is claimed in \cite{arute2019quantum} by demonstrating the sampling task on 53 qubits with 20 depths in two dimensions. Their gate fidelity is about $0.6\%$ and the final fidelity of the computation is 0.002. The similar experiments results are further confirmed on 56 qubits \cite{PhysRevLett.127.180501} and 60 qubits \cite{zhu2021quantum} with final fidelity 0.0007 and 0.0004, respectively.

However, whether the experiment results (with noisy devices) can be simulated with the current classical computer is under debate. Particularly, the recent work \cite{gao2021limitations} utilize the shortcoming of metric of fidelity i.e. linear cross-entropy benchmark (XEB)  to achieves a comparable fidelity with experiments within just a few seconds by a classical algorithm.

Tensor network methods are the most used algorithms for simulating RQCs \cite{villalonga_flexible_2019,Hyper2021optimized,Alibaba2020,PhysRevLett.128.030501,pan2021solving,PhysRevX.10.041038,gao2021limitations}. The amplitude of any basis can be obtained by contracting its corresponding tenser network. However, the optimal contraction of tensor network usually relies on heuristic algorithms \cite{Hyper2021optimized}, and the exact scaling of the simulation is unclear. Furthermore, instead of simulating real noisy quantum devices, these classical simulations primarily aim to achieve higher linear XEB values. How the computational power of the noisy quantum devices scale with qubits and noise rate is very important for understanding the
limitation and utility of quantum devices in NISQ era \cite{Noh2020efficientclassical,PhysRevA.104.022407}. Recent work \cite{Noh2020efficientclassical} focused on studying the RQCs with real physical noise in one dimension, which shows that adding more qubits doesn't lead to an exponential increase of simulation cost due to area law scaling of entanglement in one dimension. However, the understanding for computational power of noisy RQC in two dimension which is relevant to experiment setting is
still lacking.

Here, we study the RQCs with depolarizing noise on two-dimensional square lattices and use the operator entanglement entropy (EE) \cite{PhysRevA.63.040304,PhysRevA.76.032316,PhysRevA.78.022103}, which quantifies the number of parameters to faithfully represent the density matrix state using tensor network, as the indicator of the simulation cost. For calculating the operator EE easily, we use  matrix product operators (MPO) method. Our numerical results show that the maximum achievable operator EE of RQCs with constant noisy  change with system size according to the area law in two dimensions. More specifically, it grows linearly with the side length of the lattice instead of the number of qubits. In addition, our results also show that the maximum achievable operator EE is a power function of the reciprocal of noise rate in a given 2D system. We utilize the entropic inequalities to analyze the dynamics of operator EE, which indicates that the area law scaling holds for any
non-zero constant noise rate, and the volume law scaling can be achieved only if the noise rate decreases as a power function with the side length of the system.

\section{Noisy random quantum circuits in two dimensions}
	
We consider a $n = L_{1}\times L_{2}$ ($L_{1}\le L_{2}$) qubits layout on square  lattice shown in Fig.\ref{fig1}(a), and use the circuit architecture proposed in \cite{boixo2018characterizing,markov2018quantum} to generate RQCs with these qubits. According to \cite{boixo2018characterizing,markov2018quantum}, the two-qubit gates are implemented only between the nearest-neighbor sites with the order shown in Fig.\ref{qc} and each two-qubit gate is set as a two-qubit Haar-random gate; see details in Appendix \ref{rct}. The depth of the circuit is the number of the two-qubit-gate layers applied. The measurement outcomes bit string $x\in\{0,1\}^{n}$ in a computational basis has probability $P_{C}(x)$. The probability $P_{C}(x)$ will eventually converge to the Porter-Thomas distribution \cite{PhysRev.104.483,boixo2018characterizing}.
 For 2D lattices, the required depth for convergence to the Porter-Thomas distribution is proportional to $\sqrt{n}$ \cite{boixo2018characterizing}.

To simulate the effect of the noise in the RQCs concretely, we suppose each two-qubit gate is corrupted by a depolarizing noise
channel independently \cite{doi:10.1126/science.1145699}. Therefore, the effect of each noisy two-qubit gate can be simulated with the following channel :
\begin{equation}\label{noi_ga}
	\mathcal{N}(\rho) = (\mathcal{E} \circ \mathcal{U}) (\rho),
\end{equation}
where $\mathcal{U}(\rho)=U\rho U^\dag$ ($U$ is the Haar-random two-qubit gate) and $\mathcal{E}(\rho)$ is the
two-qubit depolarization channel, i.e.,
\begin{equation}\label{channel}
	\mathcal{E}(\rho) = (1-p) \rho + \frac{p}{15}\sum_{E \in \mathcal{P}} E \rho E^\dag,
\end{equation}
where $\mathcal{P}=\{I,X,Y,Z\}^{\otimes2}-\{I\otimes I\}$ is 15 possible combinations of the products of two Pauli operators and $p$ is the noise rate. The tensor network representation of the noisy two-qubit gate Eq.(\ref{noi_ga}) is shown in Fig.\ref{fig1}(d). If noise sufficiently spread in the circuit, the probability distribution $P_{C}(x)$ of noisy RQCs will converges to the uniform distribution \cite{aharonov1996limitations,gao2018efficient}.

\section{MPO SIMULATION METHOD and Operator entanglement entropy}

\begin{figure}[tb]
	\includegraphics[scale=0.7]{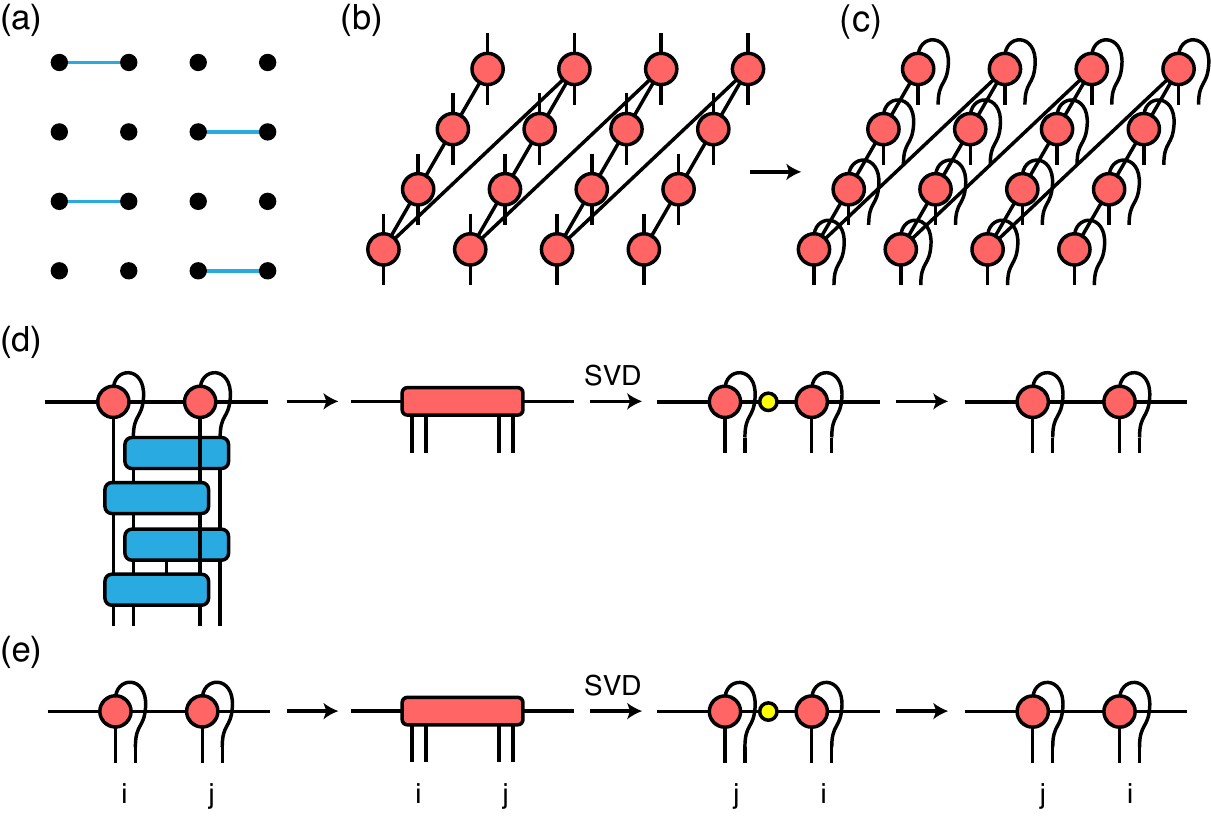}
	\caption{(a) Qubits layout on a square lattice, two-qubit gates (blue links) are applied only between the nearest-neighbor sites. (b) MPO represents qubits array in two dimensions. (c) Vectorized form of the density operator. (d) Applying a two-qubit noisy gate to an MPO, first contract the applied tensors with the two-qubit noisy gate, then decompose the resulting tensor by SVD and truncate the inner bond dimension to keep $\chi$ largest singular values. (e) Swap operation of two adjacent sites of an MPO, first contract the two tensors, then swap their physical indices by SVD and  keep the largest $\chi$ singular values.
			}
	\label{fig1}
\end{figure}

The evolution of the noisy RQCs can be regarded as the dynamics of open quantum many-body systems. A common approach to simulate the open quantum many-body systems is to use the tensor network methods \cite{RevModPhys.93.015008}. In this work, we use MPO with bond dimension $\chi$ and physical dimension $d$ as an ansatz to simulate the 2D quantum circuit, since any 2D quantum circuit can be transformed into a 1D MPO representation as Fig.\ref{fig1}(b). Further using the Choi-Jamiołkowski isomorphism, any density operator can be vectorized as:
\begin{equation}
	\sum_{ij}\rho_{ij}\left|i\right\rangle \left\langle i\right|\rightarrow\left|\rho\right\rangle\rangle \equiv\sum_{ij}\rho_{ij}\left|i\right\rangle \left|j\right\rangle.
\end{equation}
Thus, the MPO ansatz can be regarded as a matrix product state (MPS) of physical dimension $d^{2}$; see Fig.\ref{fig1}(c). Consequently, we can use
the time evolving block decimation (TEBD) like algorithms to simulate the evolution of the noisy RQCs \cite{PhysRevLett.91.147902,PhysRevLett.93.040502,PhysRevLett.93.207205}.

As a consequence of transforming the quantum circuit on 2D to 1D MPO, the local interactions between the neighbor columns become long-ranged interactions. Therefore, we need to update both local and long-ranged noisy gates. To perform local two-qubit noisy gates, the MPO which is vectorized to MPS should first transform into the canonical form around the applied qubits. Then the noisy two-qubit gate $\mathcal{N}$ is applied to the MPO which enlarged the bond dimension. Finally, a singular value decomposition (SVD) and truncation (keep $\chi$ largest singular values) are applied to restore the original MPO form. This procedure is illustrated in Fig.\ref{fig1}(d). For the long-ranged noisy two-qubit  gates, we can move one qubit to the other one through a series of local swap operations. Then the previous local update is performed. Finally, the qubit will move back to its original position by local swap. The local swap operation of two adjacent physical indices is illustrated in Figure~\ref{fig1}(e).

The time cost of TEBD simulation algorithm is $O(nt\chi^{3})$, where $t$ is the circuit depth. How large bond dimensions we should use to faithful represent a state is relevant with the degree of entanglement
in the system. A slightly entangled quantum state can be efficiently represented by an MPS  with small $\chi$ \cite{PhysRevB.73.094423}. More precisely, for 1D quantum state of length $n$, if the R\'{e}nyi entropy $S_{\alpha}(\rho_{R})=\text{log}_{2}(\text{Tr} \rho_{R}^{\alpha})/(1-\alpha)$ of a reduced density matrix $\rho_{R}$ obey $S_{\alpha}(\rho_{R}) = O(\text{log}n)$ for any bipartition $R$ and its complement $R^{c}$, then we can use an MPS with $\chi=\text{poly}(n)$ to faithful represent the state. For those states, a quantum computer may not supply advantages with respect to the classical ones. However, if $S_{\alpha}(\rho_{R}) \propto n$, an MPS with  $\chi \propto \exp(n)$ is necessary to represent the state. Since any MPO can be viewed an MPS with physical dimension $d^{2}$, the operator EE:
\begin{equation}
	S(\rho_{R})=-\text{Tr}\rho_{R}\text{log}_{2}\rho_{R},
\end{equation}
where $\rho_{R}=\text{Tr}_{R^{c}}(|\rho\rangle\rangle\langle\langle\rho|)/\text{Tr}(|\rho\rangle\rangle\langle\langle\rho|)$ for different bipartition of columns $R=1,...,L, R^{c}=L+1....,L_{2}$ of 2D $L_{1}\times L_{2}$ qubits, plays the similar role of entanglement in MPS. This quantity vanishes for maximally mixed state.

In this work, we choose the maximum over all bipartitions of columns, i.e.
\begin{equation}
	S(\rho)=\max_{R}S(\rho_{R}).
\end{equation}
as the operator EE for a certain depth. The operator EEs for a system are averaged over 40 random circuit instances for all the cases except the $7\times 7$ systems, whose operator EEs are averaged over 20 random circuit instances. For convenience, similar with \cite{Noh2020efficientclassical,PhysRevA.104.022407}, we use the trace of the density matrix $\text{Tr}(\rho)$ ($\rho$ is obtained from MPO) to measure the simulation error. The consistency between $\text{Tr}(\rho)$ and fidelity $\mathcal{F}(\rho,\sigma)=\textrm{Tr}\sqrt{\sqrt{\rho}\sigma\sqrt{\rho}}$ ($\sigma$ is the exact output density matrix) is discussed in Appendix \ref{rela}. In our simulation, the bond dimension $\chi$ used is $1000\leq\chi\leq6000$. The average $\text{Tr}(\rho)$ of almost all systems can achieve $99\%$ except several large systems whose average $\text{Tr}(\rho)$ also achieves $98\%$; see more details in Appendix \ref{rela}. In addition, the spectrum of singular values of output states always decrease exponentially fast, such that the effect of cutting bond dimension can be quite small when $\chi$ is sufficiently large; more details are discussed in Appendix \ref{sv_de}.

\section{Numerical results}

\begin{figure}[tb]
	\includegraphics[scale=0.5]{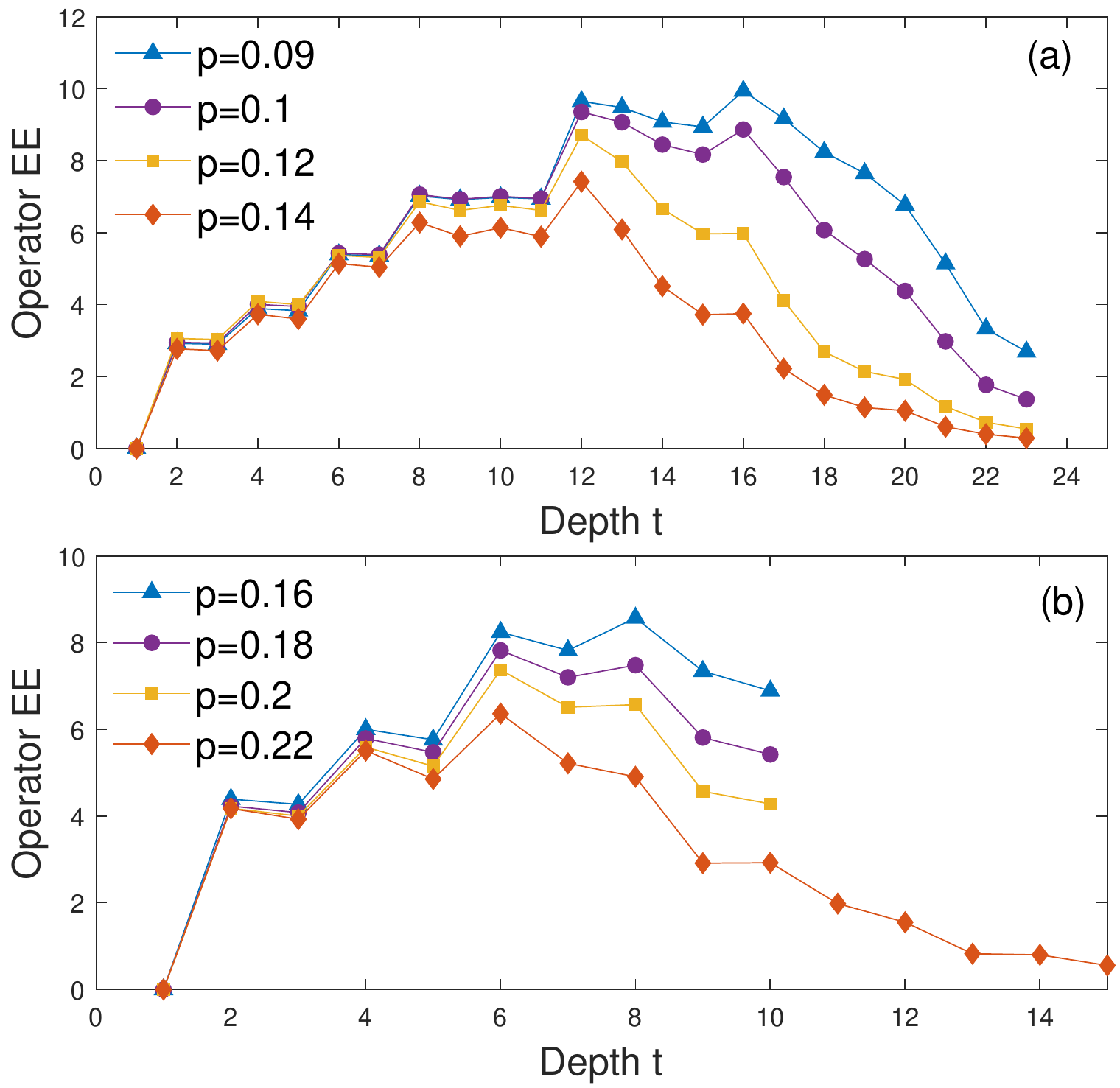}
	\caption{Operator entanglement entropy as a function of circuit depth $t$ on (a) $4\times 4$ systems with $0.09\leq p\leq0.14$ and (b)  $6\times 6$ systems with $0.16\leq p\leq0.22$
	}
	\label{fig2}
\end{figure}

Firstly, we calculate the operator EE $S(\rho)$ as functions of the circuit depth $t$ and noise rate $p$ with fixed system sizes.  Fig.\ref{fig2}(a) and (b) depict the results of $4\times 4$ and $6\times 6$ systems with different error rate $p$, respectively (the error bars are too small  to  display, therefore we don't include them here). In all of the cases, the operator EEs first increase with the depth of the circuit, which indicates the entanglement built by the two-qubit Haar-random gates is dominant at this region. After achieving the maximum at a certain circuit depth, i.e. peak circuit depth, operator EEs gradually decay as the circuit depth increases, indicating that the noise is dominant at this period and corrupts the entanglement of the states. The system continually lost information and eventually converges to the maximally mixed states; also see Appendix \ref{sec_renyi} for second R\'{e}nyi entropy. For a given system size, the
maximal operator EE and  peak circuit depth, which determines the simulation cost, increase with the decrease of $p$, which indicates that for smaller noise rate we need larger
bond dimensions to simulate the noisy RQCs (consistent with the intractable of ideal random circuit sampling problem with classical computer).

\begin{figure}[tb]
	\includegraphics[scale=0.34]{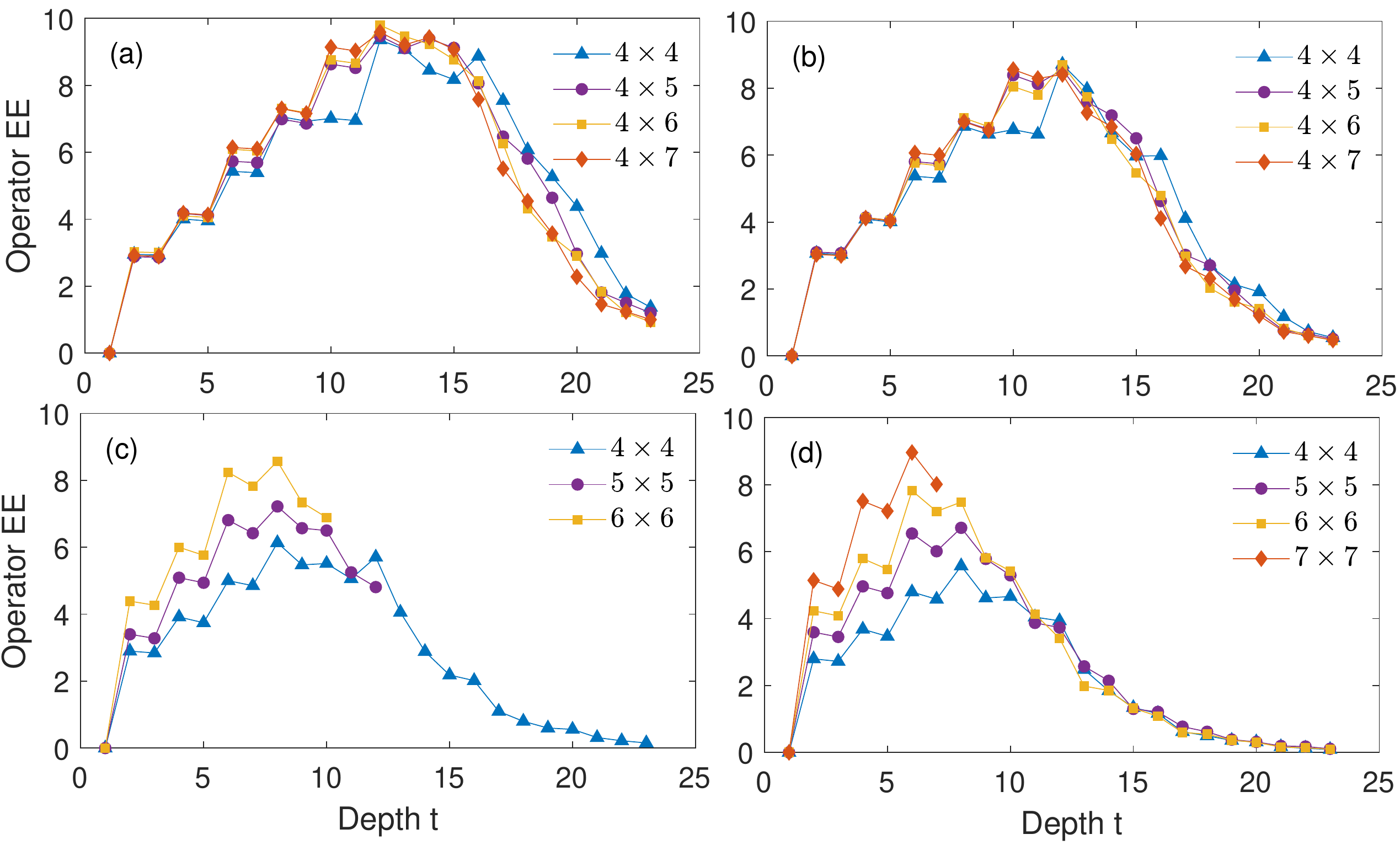}
	\caption{Operator entanglement entropy as a function of circuit depth $t$ on $4\times L$ systems for (a) $p = 0.1$ and (b) $p = 0.12$. Operator entanglement entropy as a function of circuit depth $t$ on $L\times L$ systems for (c) $p = 0.16$ and (d) $p = 0.18$.
	}
	\label{fix_e_d_si}
\end{figure}

Secondly, we explore the operator EE as functions of the circuit depth $t$ for different system sizes with a fixed noise rate $p$. We first fix the size of one side of the lattice, and change the other side. For the $4\times L$ ($L$=4, 5, 6, 7) systems shown in Fig.\ref{fix_e_d_si}(a) and (b), we calculate operator EE for $p = 0.1$ and 0.12 respectively. For $p=0.1$, operator EE achieve almost the same maximal value at the same depth $t=12$ for different $L=4,5,6,7$. For $p=0.12$, the maximal operator EE is also achieved at the almost same depth ($t=12$ for $L=4, 5, 6$ and $t$=10 for $L$=7) with the same value. However, the maximal operator EE of $p=0.12$ is smaller than the case of $p=0.1$. The results suggest that simulating the system enlarged by increasing one side of the lattice does not require simulation costs to increase exponentially.

Then we change both sides of the lattice. The operator EE results of the  $L\times L$ ($L=4,5,6,7$) systems are shown in Fig.\ref{fix_e_d_si}(c) and (d) for $p=0.16$ and $p=0.18$ respectively. As we can see, maximum achievable operator EE increases with $L$, which is different from the one-side changing situation. However, all of the values of the maximal operator EE with different system size are achieved at the almost same depth $t=8$ for $p=0.16$ and $t=$6 or 8 for $p=0.18$, respectively. In addition, the value of the  peak circuit depth in Fig.\ref{fix_e_d_si}(c) and (d) (with $p=0.16$ and $p=0.18$) are smaller than the values in Fig.\ref{fix_e_d_si}(a) and (b) with $p = 0.1$ and $0.12$. These results indicate that simulating the larger system generated by enlarging both sides need more cost.

\begin{figure}[tb]
	\includegraphics[scale=0.5]{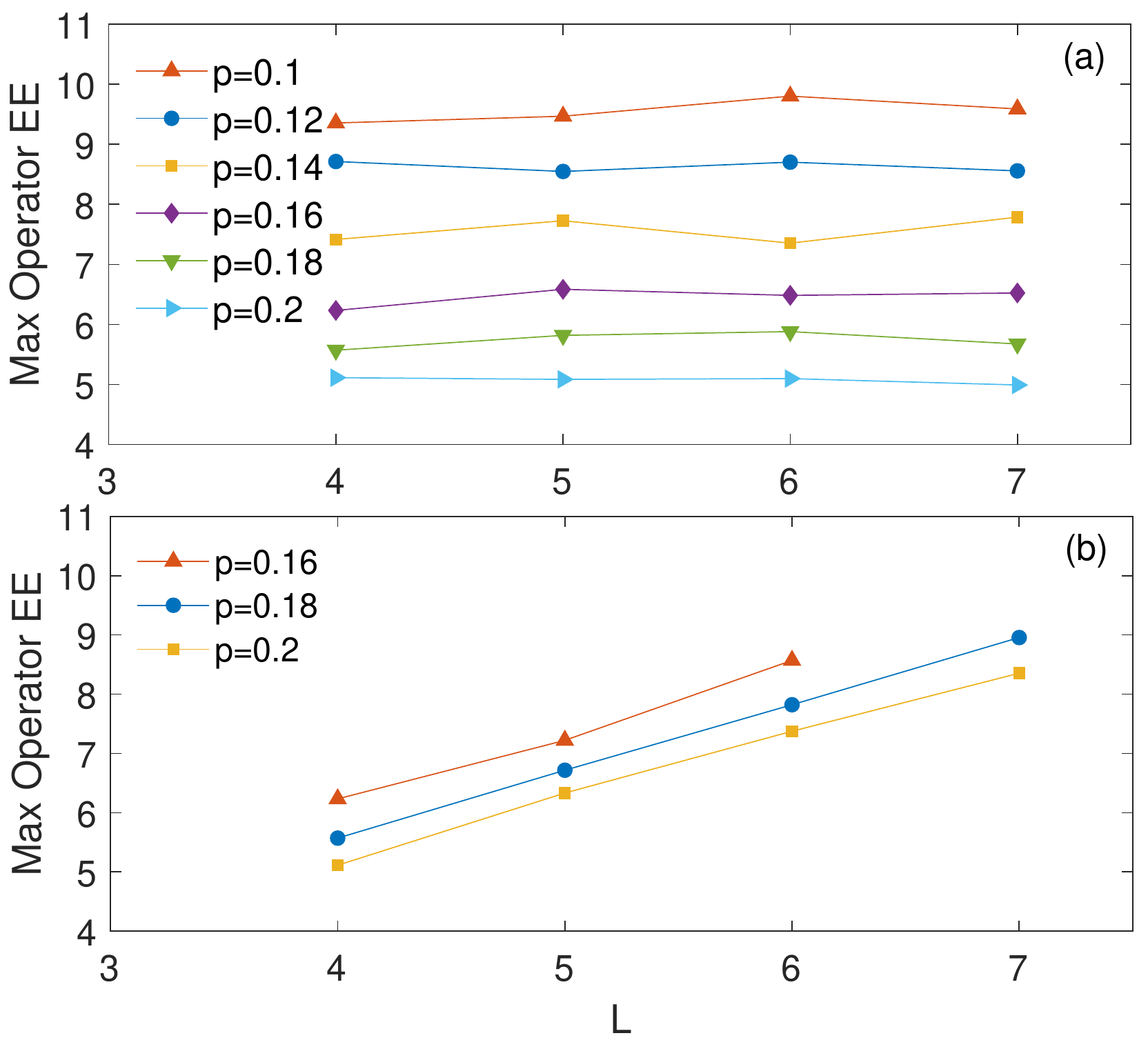}
	\caption{(a) Maximum achievable operator entanglement entropy as a function of $L$ on $4\times L$ systems for various noise rate $0.1\leq p\leq0.2$. (b) Maximum achievable operator entanglement entropy as a function of $L$ on $L\times L$ systems for various noise rate $0.16\leq p\leq0.2$.
	}
	\label{area_law}
\end{figure}

To explicitly demonstrate the relation between the maximum achievable operator EE $S_{\max}$ with the system size under constant error rate, we collect $S_{\max}$ obtained similarly as in Fig.\ref{fix_e_d_si}. The  $S_{\max}$ of $4\times L$ ($L=4,5,6,7$) systems with noise rate $0.1\leq p\leq0.2$ are shown in Fig.\ref{area_law}(a). As we can see, the $S_{\max}$ is independent of $L$ for all error rate $p$, which indicates the simulation cost only increases polynomially with system size, which is consistent with the conclusion of 1D systems in Ref.~\cite{Noh2020efficientclassical}.  However, there are some differences between our results and the ones of Ref.~\cite{Noh2020efficientclassical}. In one dimension, the $S_{\max}$ first increases linearly with the system size until it reaches certain characteristic system size, above which the $S_{\max}$ converges to a constant. For our 2D $4\times L$ systems, no such characteristic system size is found, and
the reason will be explained in the following section. The $S_{\max}$ of the $L\times L$ ($L=4,5,6,7$) systems are plotted in Fig.\ref{area_law}(b) for $0.16\leq p\leq0.2$. As we can see, the $S_{\max}$ increases linearly with side length of the system for all error rate $p$. These results suggest that the maximum achievable operator EE has area law scaling for noisy RQCs in two dimensions. It is different from  the case of the noiseless random circuits, the maximum achievable operator EE of which has volume law scaling \cite{PhysRevLett.71.1291,PhysRevLett.72.1148,PhysRevLett.77.1,Dahlsten_2007,harrow2018approximate}.

\begin{figure}[tb]
	\includegraphics[scale=0.3]{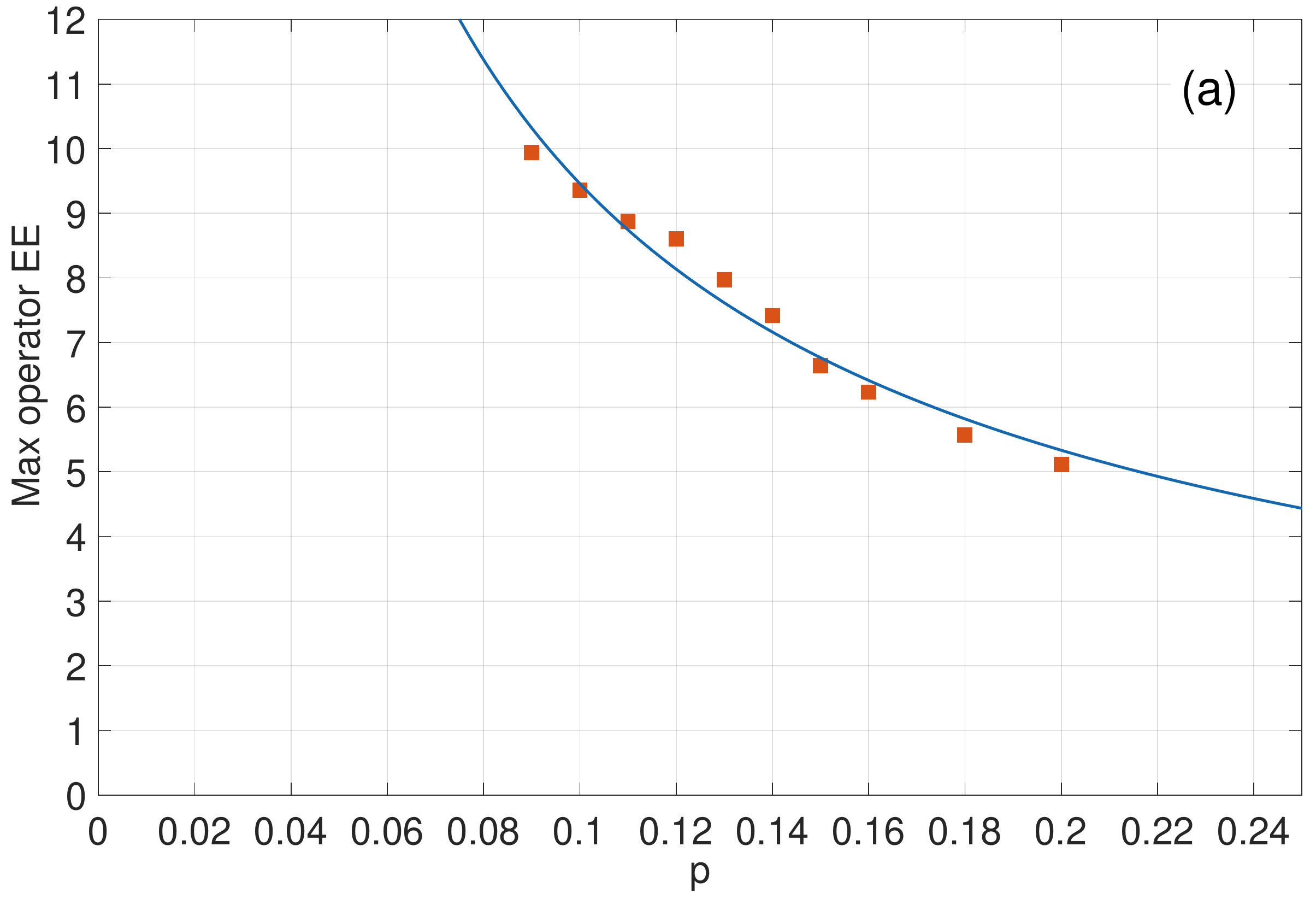}
	\includegraphics[scale=0.3]{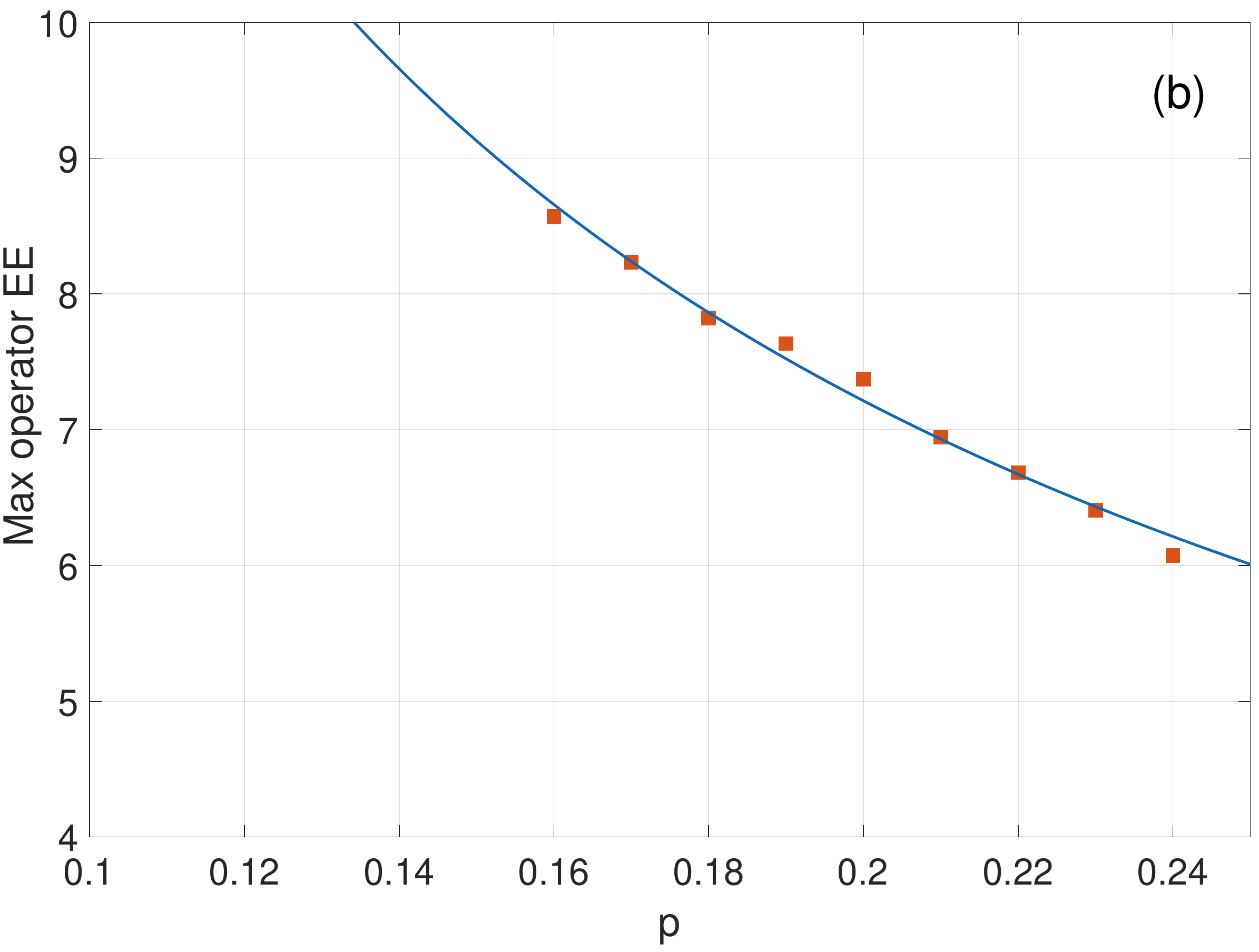}
	\caption{Maximum achievable operator entanglement entropy as a function of noise rate $p$ for (a) $4\times 4$ systems when $0.09\leq p\leq0.2$ and (b) $6\times 6$ systems when $0.16\leq p\leq0.24$. Blue lines represent power function fitting.
	}
	\label{fit}
\end{figure}

Finally, we explore how $S_{\max}$ changes with the noise rate $p$ for a given lattice system. In Fig.\ref{fit}, we show the results of the $4\times 4$ and $6\times 6$ lattice respectively, when $0.09\leq p\leq0.24$. The relation between the $S_{\max}$ and the noise rate $p$ can be well fitted by power functions $S_{\max}(p) = cp^{a}$. For $4 \times 4$ system $S_{\max}(p) = 1.41 \cdot p^{-0.8267}$, and the corresponding $95\%$ confidence bound of the estimated coefficients $a$ is $(-0.9498, -0.7035)$. For $6 \times 6$ system $S_{\max}(p) =  1.932 \cdot p^{-0.8185}$, and the corresponding $95\%$ confidence bound of the estimated coefficients $a$ is $(-0.8998, -0.7373)$. $a<0$ clear shows that $S_{\max}$ increases as the noise rate $p$ decreases. The ratio of coefficients $c$ of these two systems is $1.932/1.41 \approx 1.4$, which is close to the ratio of side length $6/4$. Therefore, combining with the results shown in Fig.\ref{area_law}, we can deduce that the maximum achievable operator EE obey
\begin{equation}\label{maxee_num}
	S_{\max}\propto L_{1}\cdot p^{-0.8}.
\end{equation}

\section{Theoretical analysis}

We can use inequalities of entanglement entropy to analyze the operator EE dynamics  of the noisy RQCs under some assumptions and explain numerical observations in the last section, especially the form ($\ref{maxee_num}$).

For a noisy chaotic quantum circuit, its output density matrix can be approximated by \cite{boixo2018characterizing}
\begin{equation}\label{noi_den}
	\rho_{\text{out}} \simeq  \alpha\left|\psi\right\rangle \left\langle \psi\right|+(1-\alpha)\frac{I}{2^{n}},
\end{equation}
where $\left|\psi\right\rangle$ is  the ideal output state. The circuit fidelity $\alpha$ can be approximated by $\alpha \simeq e^{-b_{0}pnt}$ ($b_{0}$ is a constant) \cite{PhysRevA.69.062317,boixo2018characterizing,dalzell2021random}, which was also observed in experiments \cite{barends_digital_2015,arute2019quantum,PhysRevLett.127.180501,zhu2021quantum}. In Appendix \ref{sec_renyi}, we calculate the second R\'{e}nyi entropy of the output states and find good agreement between numerical calculation and theoretical formula derived from Eq.(\ref{noi_den}), which confirm the the validity of ansatz Eq.(\ref{noi_den}) and the form of fidelity $\alpha$.

With the form of the output state (\ref{noi_den}) and utilizing the inequalities of entanglement entropy, the operator EE of noisy RQCs can be estimated  as
\begin{equation}\label{mpoee_1}
	\overline{S(\rho_{\text{out}} )} \simeq  \frac{2}{1+e^{2b_{0}pn(t-\frac{\text{In2}}{2b_{0}p} )}}  \overline{S^{\text{von}}(\left | \psi  \right \rangle )},
\end{equation}
where $S^{\text{von}}(\left | \psi  \right \rangle )$ is the von Neumann entropy of ideal output state for certain bipartitions and $\overline{\cdots}$ mean average over RQCs. The detailed derivation of Eq.(\ref{mpoee_1}) can be found in Appendix \ref{dev}.

The von Neumann entropy of $\left | \psi  \right \rangle$ in a perfect quantum chaotic dynamics in two dimensions is typically growing as \cite{PhysRevLett.112.011601,PhysRevX.9.021033,PhysRevX.7.031016}
\begin{equation}
	\overline{S^{\text{von}}(t)} \simeq  b_{1}L_{1}\cdot \min (b_{2}t, L_{2}),
	\label{von}
\end{equation}
where $b_{1}$ and $b_{2}$ are constants. Here we have assumed $L_{1} \le L_{2}$ and choose cut at the middle of the systems. Plugging the Eq.(\ref{von}) into Eq.(\ref{mpoee_1}) and using the step function to approximate the Fermi–Dirac distribution-like function $1/(1+e^{2b_{0}pn(t-\frac{\text{In2}}{2b_{0}p} )})$, we can obtain the maximum achievable operator EE as:
\begin{equation}\label{MPOEEMAX}
	S_{\text{max}} \simeq \left\{\begin{matrix}
		2b_{1}L_{1}L_{2},&L_{2} \leq \text{ln}2\cdot b_{2}/(2b_{0}p)\\
		\text{ln}2\cdot b_{1}b_{2} L_{1}/(b_{0}p),&L_{2} > \text{ln}2\cdot b_{2}/(2b_{0}p).
	\end{matrix}\right.
\end{equation}
with the peak circuit depth
\begin{equation}\label{optd}
	t_{\text{p}} \simeq \left\{\begin{matrix}
		L_{2}/b_{2},&L_{2} \leq \text{ln}2\cdot b_{2}/(2b_{0}p)\\
		\text{ln}2/(2b_{0}p),&L_{2} > \text{ln}2\cdot b_{2}/(2b_{0}p).
	\end{matrix}\right.
\end{equation}
See Appendix \ref{dev} for details.

With Eq.(\ref{MPOEEMAX}), for a given constant $p$, we can see that $S_{\text{max}}$ will first increase as a volume law with enlarging $L_{2}$, then become area law after transition size $L_{\text{tran}}=\text{ln}2\cdot b_{2}/(2b_{0}p)$. To keep the volume law scaling  (avoid transition) of $S_{\text{max}}$, the noise rate $p$ should scale as $O(L_{2}^{-1})$.

In principle, the theoretical analysis works for higher dimensions. As long as the entanglement dynamics of noiseless RQCs in $z$ dimension obey $\overline{S(t)} \propto L^{z-1}\min (b_{2}t, L) $ \cite{PhysRevX.7.031016} and circuit fidelity decay as $\alpha \simeq e^{-b_{0}pnt}$, the area law of operation EE of noisy RQCs will hold.

The numerical results in this work and in Ref. \cite{Noh2020efficientclassical} can be understood using Eq.(\ref{MPOEEMAX})  and Eq.(\ref{optd}).
Ref. \cite{Noh2020efficientclassical} study 1D noisy RQCs and observe that $S_{\text{max}}$ increases linearly in the system size until the system size reaches a certain characteristic size, above which $S_{\text{max}}$ is independent of the system size. These results are well described by  Eq.(\ref{MPOEEMAX}) for $L_{1}$=1. For the 2D noisy RQCs calculated in this work, when we fix the noise rate $p$, the $S_{\text{max}}$ only linearly grows with the side length as we observe in Fig.\ref{area_law}, which is consistent with the case of $L_{2} > \text{ln}2b_{2}/(2b_{0}p)$ in Eq.(\ref{MPOEEMAX}). We don't observe the volume law scaling region and the transition size, which is because, for the noise rate $p$ we used, the corresponding transition size is relatively small and can not appear in our setting ($L_{2}\ge L_{1} \ge 4$). In addition, if we fix the system size, $S_{\text{max}}$ has power law scaling with $p$, which justifies that we can use power functions to fit the results in Fig.\ref{fit}. We note that the fitting function in Fig.\ref{fit} is $S_{\max}(p)\propto p^{-0.8}$ instead of $p^{-1}$. This deviation is mainly because the concrete entanglement dynamics of ideal output and the  decay rate of fidelity depend on the architecture and spatial dimension of the quantum circuit, such that the theoretical assumption can not fully be satisfied in our numerical simulation.  In region of $L_{2} > \text{ln}2b_{2}/(2b_{0}p)$ in Eq.(\ref{optd}), the peak circuit depth $t_{p}$ is independent of the system size and relies only on the noise rate $p$, which agree with our observation in Fig.\ref{fix_e_d_si}. The $t_{p}$ increases with the decrease of noise rate, which is also observed in Fig.\ref{fig2}. Therefore, the dynamics of the operator EE of the noisy RQCs can be regarded as the result of competition between two processes: the system lost information exponentially fast and entanglement of quantum chaotic circuit ballistically spreads before achieving maximal entanglement.

\section{DISCUSSION AND CONCLUSION}

In this work, we find area law scaling behavior of operator EE for noisy RQCs in 2D experiment relevant architecture. We don’t observe any threshold of noise rate below which the scaling behavior of maximum
achievabl operator EE is volume law. Our numerical simulation results well agree with the theoretical analysis. The theoretical
analysis indicates that the area law scaling holds for any non-zero constant noise rate. In addition, the maximum achievable operator EE just obeys volume law for systems
smaller than some p-dependent transition size, above
which the maximum achievable operator
EE become area law scaling. To maintain volume
law scaling, the noise rate should decrease as a power
function with the side length of the system.

The area law scaling of the operator EE in 2D architecture hints that it is possible to efficiently represent the output density matrices of the noisy RQCs by projected entangled pair operators (PEPOs). However, it does not immediately lead to classical simulatability of noisy RQCs because exact contracting the PEPOs is a $\#p$ complete problem \cite{PhysRevLett.98.140506,PhysRevResearch.2.013010}. Nevertheless, it is also possible to efficiently contract the PEPOs approximately by many numerical algorithms used in projected entangled pair states (PEPSs) \cite{PhysRevLett.99.120601,PhysRevLett.111.090501,PhysRevB.104.235141,PhysRevB.95.195154}.

From an experimental perspective, our results suggest that only adding qubits along one side of the qubits layout may not increase the classical simulation cost too much if the system is in the area law region. Recent experimental work \cite{zhu2021quantum} fabricated a $6\times 11$ rectangular superconducting qubit array in two dimensions, which add more qubits along one side than the $6\times 9$ qubit array of work \cite{arute2019quantum}. If the qubit layout is closer to square array, it would have a higher capacity of entanglement and make classical simulation more difficult. Furthermore, if we expect to maintain volume law scaling to achieve stronger quantum advantage, noise rate $p$ need to decrease as $p=O(L^{-1.25})$ for experiment relevant architecture. Therefore, reducing the noise rate of two-qubit gates is the primary task to increase the computational power of RQCs in the NISQ era.

\acknowledgements
This work was funded by Innovation Program for
Quantum Science and Technology 2021ZD0301200 and S.D. is
supported by Chinese National Science Foundation Grant
No. 12104433.  The numerical calculations were done on the USTC HPC facilities.

\appendix
\section{Random circuits in two dimensions}\label{rct}
\begin{figure}[tb!]
	\includegraphics[scale=0.4]{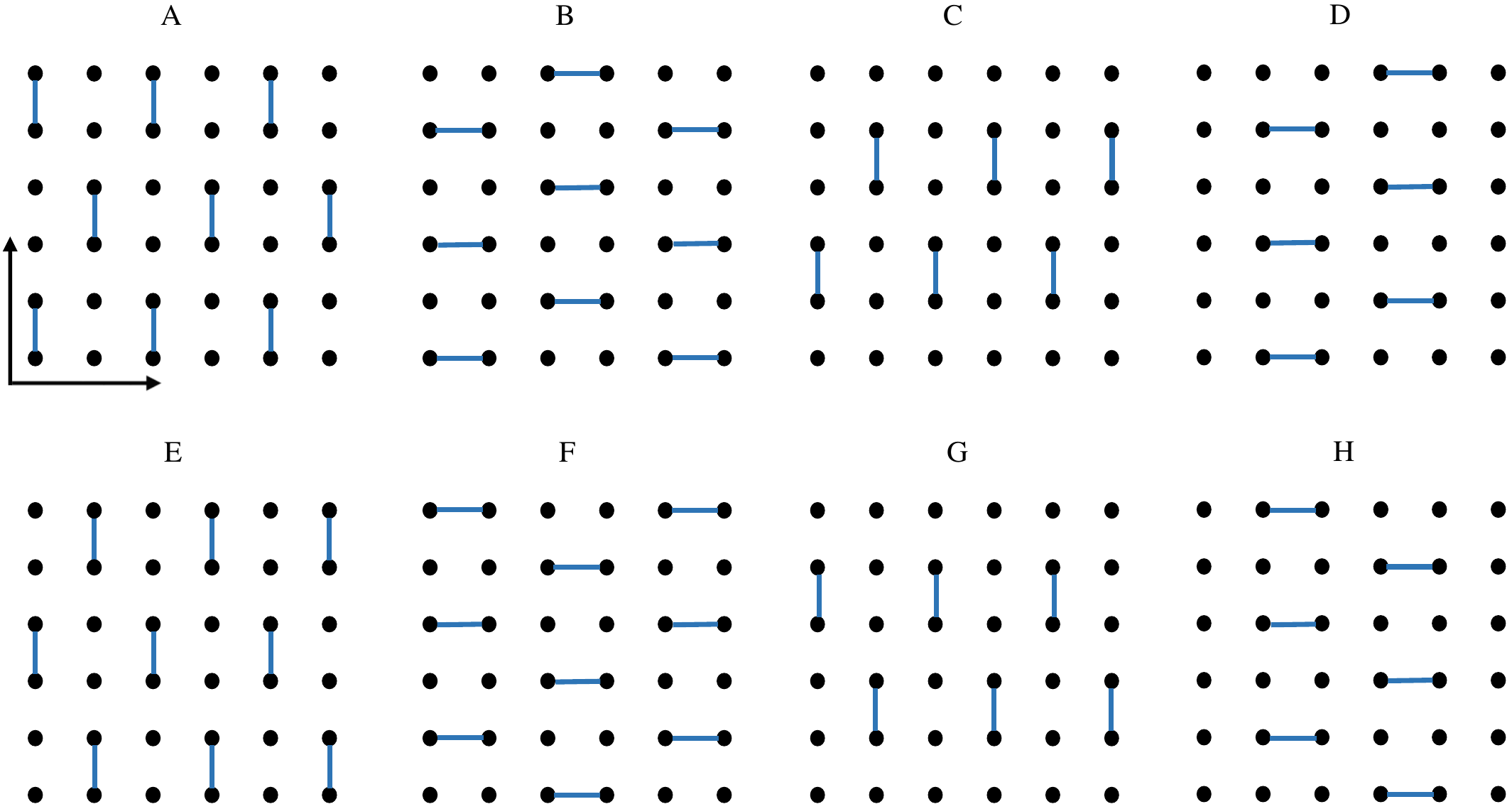}
	\caption{The architecture of random quantum circuits. The black circles represent qudits and blue links represent two-qubit gates. The layer is repeated in the sequence of ABCDEFGH.
	}
	\label{qc}
\end{figure}

The architecture of RQCs used by experimental works \cite{arute2019quantum,PhysRevLett.127.180501,zhu2021quantum} was proposed originally by work \cite{boixo2018characterizing}. Latter, work \cite{markov2018quantum} proposed a revised version to make the classical simulation more difficult. In this work, we use the revised architecture of quantum circuits. The quantum circuits are generated on 2D square grids by recycling eight different layers of two-qubit gates as shown in Fig.\ref{qc}. We use two-qubit Haar-random gates to generate an ensemble of quantum gates.

\begin{figure}[b]
	\includegraphics[scale=0.36]{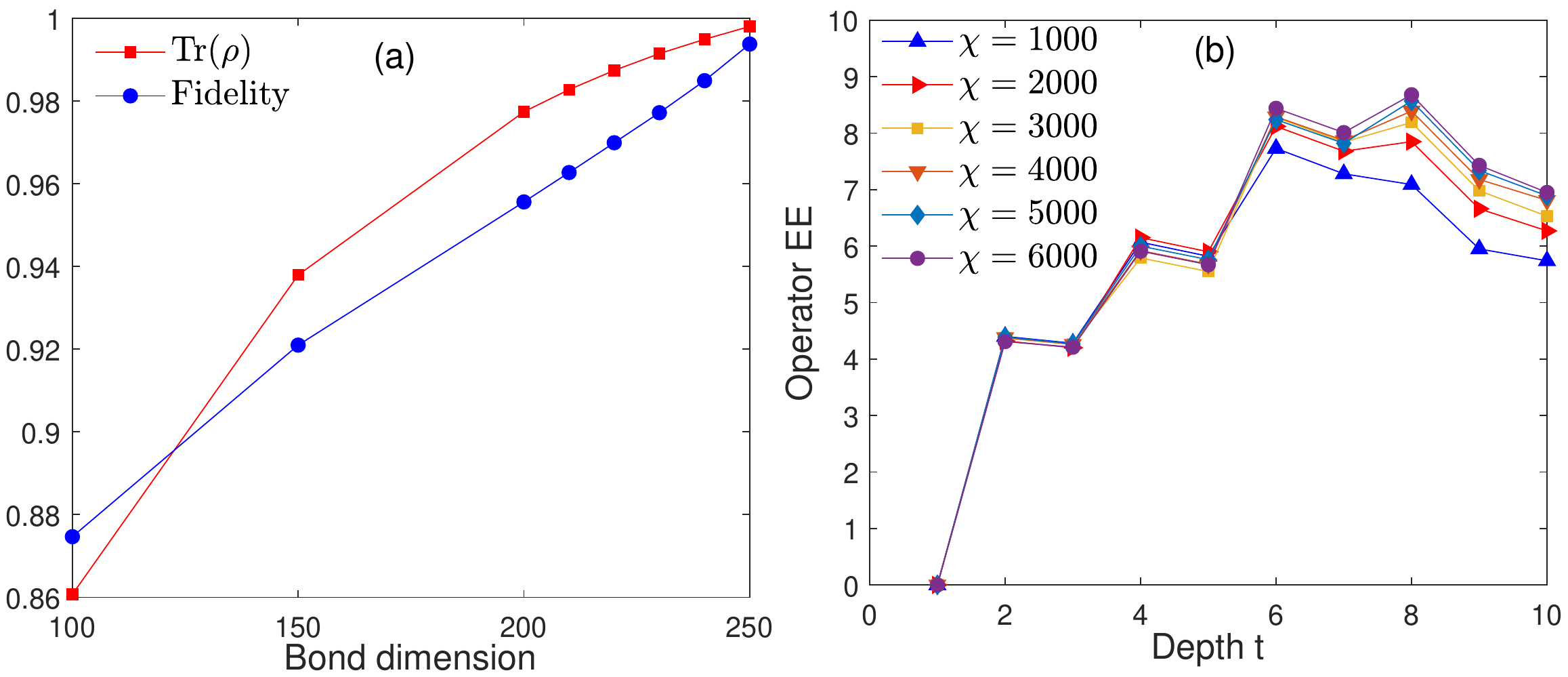}
	\caption{(a) $\text{Tr}(\rho)$ and fidelity $\mathcal{F}(\rho,\sigma)$ as a function of bond dimension $\chi$ for 2D $3 \times 3$ noisy quantum circuit with noise rate $p=0.08$, and the circuit depth is 24. (b) The operator EE  as functions of the circuit depth $t$ for $ 6 \times 6$ system with $p=0.16$ calculated by MPOs with different $\chi$.
	}
	\label{fidelity}
\end{figure}

\section{$\text{Tr}(\rho)$ as an indicator of simulation error}\label{rela}
In this work, we use $\text{Tr}(\rho)$ to measure the simulation error  caused by the truncation bond dimension $\chi$. When no truncation is introduced, $\text{Tr}(\rho)$ is unity, whereas
$\text{Tr}(\rho) < 1$ if the bond dimensions are truncated. To investigate the relation between  $\text{Tr}(\rho)$  and fidelity $\mathcal{F}(\rho,\sigma)=\textrm{Tr}\sqrt{\sqrt{\rho}\sigma\sqrt{\rho}}$, where $\sigma$ denotes the output density matrix of exact simulation, we simulate a 2D $3 \times 3$ system  with noise rate $p=0.08$ and depth 24 using MPO with different bond dimensions, and compare $\text{Tr}(\rho)$ and fidelity $\mathcal{F}(\rho,\sigma)$  for the final states. As we can see from Fig.\ref{fidelity}(a), when $\text{Tr}(\rho)$ achieve $99\%$, fidelity  can achieve $98\%$ and when  $\text{Tr}(\rho)$ achieve $98\%$ fidelity can achieve $96\%$. Therefore, $\text{Tr}(\rho)$ can be regarded as a good measure of the simulation error when the value of $1- \text{Tr}(\rho)$ is small.

In this work, we always keep $1- \text{Tr}(\rho)$ small to accurately simulate noisy quantum circuits. The average $\text{Tr}(\rho)$ of all system can achieve $99\%$ except $7 \times 7$ system with $p=0.18$ ($\text{Tr}(\rho)$ = 98.2\%), $4 \times 7$ system with $p =0.1$ ($\text{Tr}(\rho)$ = 98.2\%) and $ 6 \times 6$ system with $p=0.16$ ($\text{Tr}(\rho)$ = 98.4\%).

We also calculate the operator EE  as functions of the circuit depth $t$ for $ 6 \times 6$ system with $p=0.16$ using MPOs with different $\chi$. The operator EEs of all the cases are averaged over 40 random circuit instances except the case of $\chi =6000$, the operator EEs of which are averaged over 20 instances. As we can see in Fig.\ref{fidelity}(b), the operator EEs have converged when $\chi =5000$, the bond dimension we employed for this system.

\section{Spectrum of singular values}\label{sv_de}

\begin{figure}[tb]
	\includegraphics[scale=0.5]{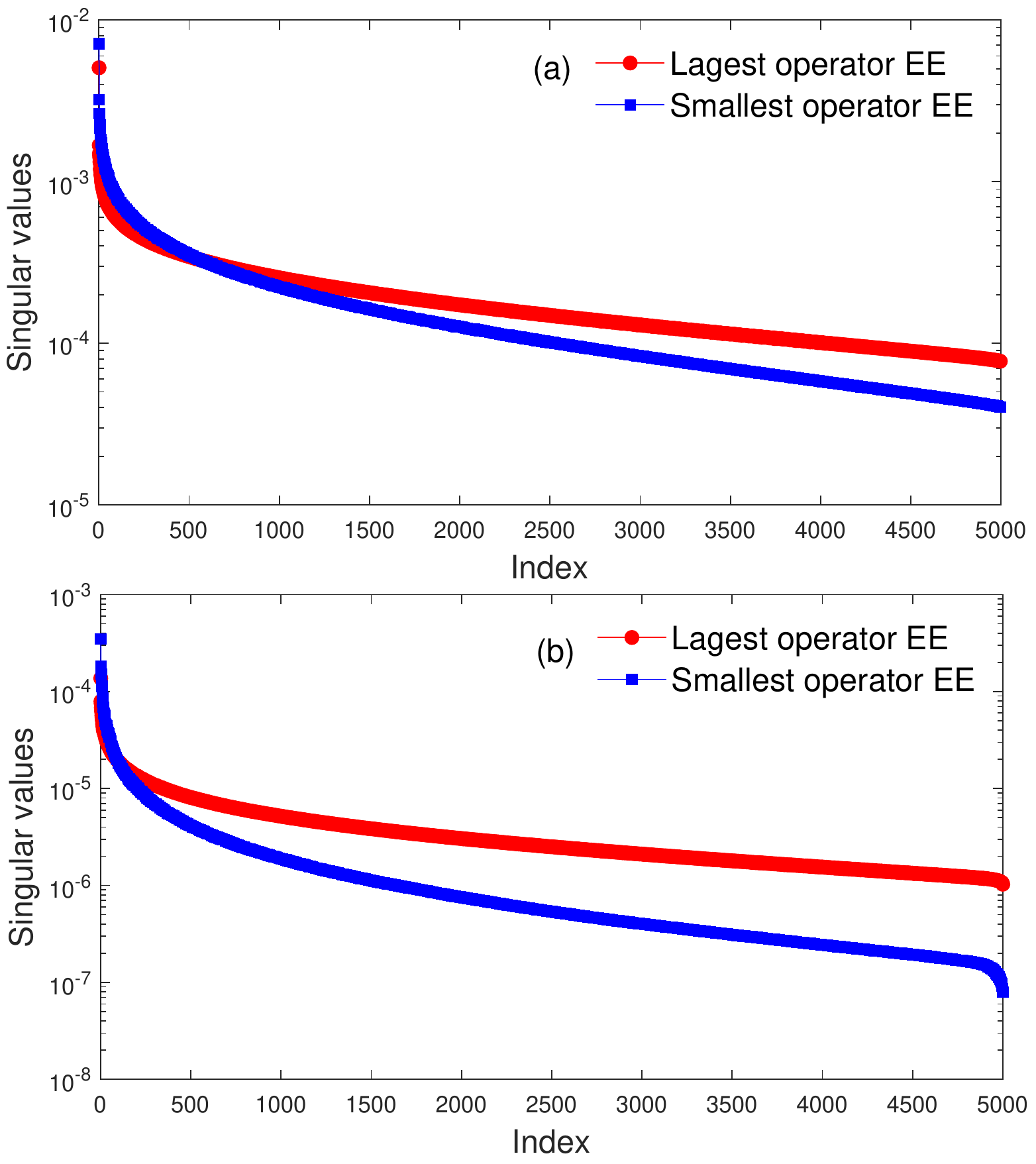}
	\caption{The spectrum of singular values in the descending order for (a) $4\times4$ systems with $p=0.09$ and (b) $6\times6$ systems with $p=0.16$ at peak circuit depth. Bipartition is chosen at the middle of the systems. The instances with largest and smallest operator EE are chosen among 40 different random circuits.
	}
	\label{sv}
\end{figure}

Figure~\ref{sv}(a) and (b) presents the spectrum of singular values for $4\times4$ systems with $p=0.09$ and $6\times6$ systems with $p=0.16$, respectively. We choose the instances with the largest and smallest operator EE among 40 instances. We choose the peak circuit depth and cut at the middle of the systems. As we can see, the singular values decrease exponentially fast for both systems. Thus, when we truncate MPO to incur a truncation error $\epsilon$  where $\epsilon$ is the sum of the squares of the discarded singular values, the needed bond dimension and simulation cost only scale as $O(\text{ln}(1/\epsilon))$ \cite{Noh2020efficientclassical,PhysRevB.73.094423,2001.00021}. Therefore, the effect of cutting bond dimension can be neglected when $\textrm{log}_{2}\chi\gg S$.

\section{Second R\'{e}nyi entropy}\label{sec_renyi}

\begin{figure}[tb]
  \centering
  \includegraphics[width=0.46\textwidth]{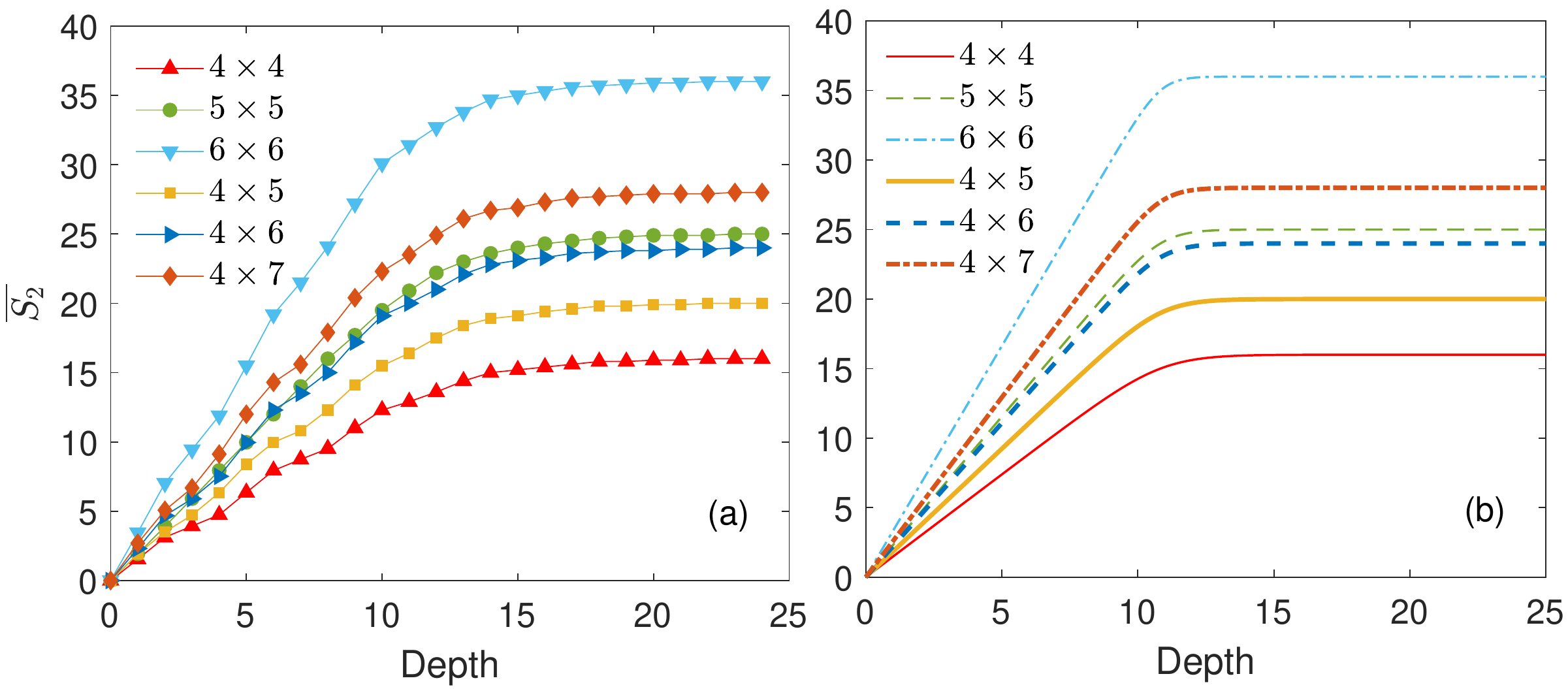}
  \caption{The second R\'{e}nyi entropy as function of depth for different system sizes
with noise rate $p=0.16$, which are obtained by (a) numerical calculation and (b) Eq.(\ref{sec_ren_th}) respectively.}
  \label{sec_renyi_fig}
\end{figure}

The second R\'{e}nyi entropy $S_{2}=-\text{log}_{2}(\text{Tr}(\rho^{2}))$ quantifies the purity of the output state. For pure states $S_{2}=0$ and for maximally mixed states $S_{2}=n$. In Fig.\ref{sec_renyi_fig}(a), we present the second R\'{e}nyi entropy as a function of depth for different system sizes
with noise rate $p=0.16$, which are obtained by numerical calculation and averaged over 40 random circuit instances. As we can see, the second R\'{e}nyi entropy increases monotonously with depth and converges to maximum $n$. The inflection point of each curve is at about depth $t=12$, which is independent of system size.

Using Eq.(\ref{noi_den}), we can also obtain the second R\'{e}nyi entropy. Firstly, $\text{Tr}(\rho^{2})=\alpha^{2} +(1-\alpha^{2})/2^{n}$. Then, if $\alpha \simeq e^{-b_{0}pnt}$, we can obtain the average second R\'{e}nyi entropy
\begin{equation}\label{sec_ren_th}
    \overline{S_{2}(t,n)}=-\text{log}_{2}(e^{-2b_{0}pnt}+\frac{1}{2^{n}}(1-e^{-2b_{0}pnt})).
\end{equation}
Fig.\ref{sec_renyi_fig}(b) plot Eq.(\ref{sec_ren_th}) as function of depth for different system sizes $n$. Here we choose constant $b_{0}=0.2$. As we can see, the curves obtained by Eq.(\ref{sec_ren_th}) are consistent with the results of numerical calculation in Fig.\ref{sec_renyi_fig}(a). The agreement on both results indicates the validity of ansatz Eq.(\ref{noi_den}) and the form of fidelity $\alpha \simeq e^{-b_{0}pnt}$.

\section{The derivation of theoretical analysis}\label{dev}
The output density matrix can be written as the convex combination of other density matrices $\rho_{\text{out}}=\sum_{i}p_{i}\rho_{i}$, the vectorized form of which is
\begin{equation}\label{vec}
	\left|\rho\right\rangle\rangle =\sum_{i}p_{i} \left|\rho_{i}\right\rangle\rangle.
\end{equation}
Therefore, the vectorized form of Eq.(\ref{noi_den}) is
\begin{equation}\label{noi_den_vec}
	\left|\rho\right\rangle\rangle =\alpha \left | \psi  \right \rangle \otimes \left | \psi  \right \rangle +(1-\alpha )| \frac{I}{2^{n}}\rangle\rangle.
\end{equation}
If we want obtain the operator EE of Eq.(\ref{noi_den_vec}) for bipartition of $R$ and $R^{c}$, we should first calculate $\text{Tr}_{R^{c}}(|\rho\rangle\rangle\langle\langle\rho|)$, where
\begin{equation}\label{dens_for_vec}
	\begin{split}
		|\rho\rangle\rangle\langle\langle\rho| = \alpha^{2} \left | \psi  \right \rangle \otimes \left | \psi  \right \rangle \left\langle\psi\right| \otimes \left\langle\psi\right| +(1-\alpha )^{2} |\frac{I}{2^{n}}\rangle\rangle\langle\langle \frac{I}{2^{n}}|\\
		+\alpha (1-\alpha )\left (\left | \psi  \right \rangle \otimes \left | \psi  \right \rangle \langle\langle \frac{I}{2^{n}}| + |\frac{I}{2^{n}}\rangle\rangle \left\langle\psi\right| \otimes \left\langle\psi\right| \right)	.
	\end{split}
\end{equation}
We can define
\begin{equation}\label{normalized}
	\begin{split}
		\rho _{0} &\equiv  |\rho\rangle\rangle\langle\langle\rho| \\
		\rho _{1} &\equiv \left | \psi  \right \rangle \otimes \left | \psi  \right \rangle \left\langle\psi\right| \otimes \left\langle\psi\right| \\
		\rho _{2} &\equiv |\frac{I}{2^{n}}\rangle\rangle\langle\langle \frac{I}{2^{n}}| \\
		\rho _{3} &\equiv \left | \psi  \right \rangle \otimes \left | \psi  \right \rangle \langle\langle \frac{I}{2^{n}}| + |\frac{I}{2^{n}}\rangle\rangle \left\langle\psi\right| \otimes \left\langle\psi\right|\\
		\rho _{i} ^{u} &\equiv  \frac{\rho _{i}}{\text{Tr}{\rho _{i}} },
	\end{split}
\end{equation}
where
\begin{equation}\label{trace}
	\begin{split}
		\text{Tr}{\rho _{0}} &= \alpha^{2} +\frac{1-\alpha^{2}}{2^{n}}\\
		\text{Tr}{\rho _{1}} &= 1\\
		\text{Tr}{\rho _{2}} &= \frac{1}{2^{n}} \\
		\text{Tr}{\rho _{3}} &= \frac{2}{2^{n}} .	
	\end{split}	
\end{equation}
Using Eq.(\ref{normalized}), we can rewrite Eq.(\ref{dens_for_vec}) as
\begin{equation}\label{nor_noi_den}
	\rho _{0}^{u} = \frac{1}{\alpha^{2} +\frac{1-\alpha^{2}}{2^{n}}} \left (\alpha^{2}\rho _{1}^{u} + \frac{(1-\alpha)^{2}}{2^{n}}\rho _{2}^{u} +\frac{2\alpha (1-\alpha)}{2^{n}}\rho _{3}^{u}\right ).
\end{equation}
We can define the coefficients of Eq.(\ref{nor_noi_den}) as
\begin{equation}
	\begin{split}
		c_{1} &= \frac{\alpha^{2}}{\alpha^{2} +\frac{1-\alpha^{2}}{2^{n}}}\\
		c_{2} &= \frac{\frac{(1-\alpha)^{2}}{2^{n}}}{\alpha^{2} +\frac{1-\alpha^{2}}{2^{n}}}\\
		c_{3} &= \frac{\frac{2\alpha (1-\alpha)}{2^{n}}}{\alpha^{2} +\frac{1-\alpha^{2}}{2^{n}}} .	
	\end{split}	
\end{equation}
The reduced density matrix of $\rho _{0}^{u}$ for bipartition of $R$ and $R^{c}$ is
\begin{equation}\label{redu_sum}
	\rho_{0,R} ^{u} = \sum_{i=1}^{3}c_{i}\rho_{i,R} ^{u},  \quad \rho _{i,R} ^{u} \equiv  \text{Tr}_{R^{c}}(\rho _{i}^{u}).
\end{equation}
Then we consider to bound the entropy of $\rho_{0,R} ^{u}$ by utilizing entropic inequalities \cite{nielsen2000},
\begin{equation}\label{en_in}
	\begin{split}
		S(\rho_{\text{out}}) &\le \sum_{i}p_{i}S(\rho_{i})+H(p_{i})\\
		S(\rho_{\text{out}}) &\ge \sum_{i}p_{i}S(\rho_{i}),
	\end{split}
\end{equation}
where $H(p_{i})=-\sum_{i} p_{i}\text{log}_{2}p_{i}$ is Shannon entropy.

Using tensor network representations, $\rho _{2}^{u}$ can be visualized as
\begin{equation}
	\rho _{2}^{u} \propto |I_{R}\otimes I_{R^{c}} \rangle \langle I_{R}\otimes I_{R^{c}}| =
	\adjustbox{valign=M}{
	\begin{tikzpicture}
	\begin{scope}
		\draw (-0.3,0.9) .. controls (-0.2,-0.1) and  (0.2,-0.1) .. (0.3,1.1) ;
		\draw (-0.3,-1.1) .. controls (-0.2,0.1) and  (0.2,0.1) .. (0.3,-0.9) ;
	\end{scope}
	\begin{scope}[xshift=1cm]
		\draw (-0.3,0.9) .. controls (-0.2,-0.1) and  (0.2,-0.1) .. (0.3,1.1) ;
		\draw (-0.3,-1.1) .. controls (-0.2,0.1) and  (0.2,0.1) .. (0.3,-0.9) ;
	\end{scope}
	\end{tikzpicture}}\ .
\end{equation}
Therefore, the reduced density matrix $\rho _{2,R} ^{u}$ is
\begin{equation}\label{dig_rhr2}
	\rho _{2,R} ^{u} \propto
	\adjustbox{valign=M}{
	\begin{tikzpicture}
	\begin{scope}
		\draw (-0.3,0.9) .. controls (-0.2,-0.1) and  (0.2,-0.1) .. (0.3,1.1) ;
		\draw (-0.3,-1.1) .. controls (-0.2,0.1) and  (0.2,0.1) .. (0.3,-0.9) ;
	\end{scope}
	\begin{scope}[xshift=0.9cm]
		\draw (0,0) circle [x radius = 0.3, y radius = 0.6] ;
	\end{scope}
	\end{tikzpicture}}\ .
\end{equation}
$\rho _{3}^{u}$ can be visualized as
\begin{equation}\label{dig_rh3}
	\rho _{3} ^{u} \propto
	\adjustbox{valign=M}{
	\begin{tikzpicture}[radius=0.12cm,fill=red!60]
	\begin{scope}
		\draw(-0.3,0.9) -- (-0.3,0.2)  -- (0.7,0.2)  -- (0.7,0.9) ;
		\filldraw (-0.3,0.2) circle;
		\filldraw (0.7,0.2) circle;
	\end{scope}
	\begin{scope}[yshift=0.2cm,xshift = 0.6cm]
		\draw(-0.3,0.9) -- (-0.3,0.2) -- (0.7,0.2) -- (0.7,0.9) ;
		\filldraw (-0.3,0.2) circle;
		\filldraw (0.7,0.2) circle;
	\end{scope}
	\begin{scope}
		\draw (-0.3,-1.1) .. controls (-0.2,0.1) and  (0.2,0.1) .. (0.3,-0.9) ;
	\end{scope}
	\begin{scope}[xshift=1cm]
		\draw (-0.3,-1.1) .. controls (-0.2,0.1) and  (0.2,0.1) .. (0.3,-0.9) ;
	\end{scope}
	\end{tikzpicture}}
	\ +\
	\adjustbox{valign=M}{
	\begin{tikzpicture}[radius=0.12cm,fill=red!60]
	\begin{scope}[yshift=-0.2cm]
		\draw(-0.3,-0.9) -- (-0.3,-0.2)  -- (0.7,-0.2)  -- (0.7,-0.9) ;
		\filldraw (-0.3,-0.2) circle;
		\filldraw (0.7,-0.2) circle;
	\end{scope}
	\begin{scope}[xshift = 0.6cm]
		\draw(-0.3,-0.9) -- (-0.3,-0.2)  -- (0.7,-0.2)  -- (0.7,-0.9) ;
		\filldraw (-0.3,-0.2) circle;
		\filldraw (0.7,-0.2) circle;
	\end{scope}
	\begin{scope}
		\draw (-0.3,0.9) .. controls (-0.2,-0.1) and  (0.2,-0.1) .. (0.3,1.1) ;
	\end{scope}
	\begin{scope}[xshift=1cm]
		\draw (-0.3,0.9) .. controls (-0.2,-0.1) and  (0.2,-0.1) .. (0.3,1.1) ;
	\end{scope}
	\end{tikzpicture}}\ .
\end{equation}
Therefore, $\rho _{3,R} ^{u}$ is
\begin{equation}\label{dig_rhr3}
     \rho _{3,R} ^{u} \propto
     \adjustbox{raise=-7ex}{
     \begin{tikzpicture}[radius=0.12cm,fill=red!60]
         \begin{scope}
              \draw(-0.3,0.9) -- (-0.3,0.2)  -- (0.7,0.2)  .. controls (0.8,1.4) and  (1.2,1.4) .. (1.3,0.4) ;
              \filldraw (-0.3,0.2) circle;
              \filldraw (0.7,0.2) circle;
        \end{scope}
        \begin{scope}[yshift=0.2cm,xshift = 0.6cm]
              \draw(-0.3,0.9) -- (-0.3,0.2) -- (0.7,0.2) ;
              \filldraw (-0.3,0.2) circle;
              \filldraw (0.7,0.2) circle;
         \end{scope}
         \begin{scope}
            \draw (-0.3,-1.1) .. controls (-0.2,0.1) and  (0.2,0.1) .. (0.3,-0.9) ;
         \end{scope}
    \end{tikzpicture}}
    \ +\
    \adjustbox{raise=-9ex}{
    \begin{tikzpicture}[radius=0.12cm,fill=red!60]
         \begin{scope}[yshift=-0.2cm]
              \draw (-0.3,-0.9) -- (-0.3,-0.2)  -- (0.7,-0.2) .. controls (0.8,-1.2) and  (1.2,-1.2) .. (1.3,0);
              \filldraw (-0.3,-0.2) circle;
              \filldraw (0.7,-0.2) circle;
         \end{scope}
         \begin{scope}[xshift = 0.6cm]
              \draw (-0.3,-0.9) -- (-0.3,-0.2)  -- (0.7,-0.2) ;
              \filldraw (-0.3,-0.2) circle;
              \filldraw (0.7,-0.2) circle;
         \end{scope}
         \begin{scope}
          \draw (-0.3,0.9) .. controls (-0.2,-0.1) and  (0.2,-0.1) .. (0.3,1.1) ;
         \end{scope}
     \end{tikzpicture}}\ .
\end{equation}
We can consider to add the following pure state $\rho_{c}$ to both side of Eq.(\ref{redu_sum}),
\begin{equation}
    \rho_{c} \equiv  \frac{\alpha^{2}}{2^{\frac{n}{2}}\text{Tr}{\rho _{0}}}
    \adjustbox{raise=-9ex}{
    \begin{tikzpicture}[radius=0.12cm,fill=red!60]
         \begin{scope}
              \draw(-0.3,0.9) -- (-0.3,0.2)  -- (0.7,0.2)  .. controls (0.8,1.4) and  (1.2,1.4) .. (1.3,0.4) ;
              \filldraw (-0.3,0.2) circle;
              \filldraw (0.7,0.2) circle;
        \end{scope}
        \begin{scope}[yshift=0.2cm,xshift = 0.6cm]
              \draw(-0.3,0.9) -- (-0.3,0.2) -- (0.7,0.2) ;
              \filldraw (-0.3,0.2) circle;
              \filldraw (0.7,0.2) circle;
         \end{scope}
         \begin{scope}[yshift=-0.2cm]
              \draw (-0.3,-0.9) -- (-0.3,-0.2)  -- (0.7,-0.2) .. controls (0.8,-1.2) and  (1.2,-1.2) .. (1.3,0);
              \filldraw (-0.3,-0.2) circle;
              \filldraw (0.7,-0.2) circle;
         \end{scope}
         \begin{scope}[xshift = 0.6cm]
              \draw (-0.3,-0.9) -- (-0.3,-0.2)  -- (0.7,-0.2) ;
              \filldraw (-0.3,-0.2) circle;
              \filldraw (0.7,-0.2) circle;
         \end{scope}
     \end{tikzpicture}}.
\end{equation}
Then $c_{2}\rho _{2,R} ^{u}+c_{3}\rho _{3,R} ^{u}+\rho_{c}$ will be a pure state, the entropy of which is 0.

Therefore, utilizing Eq.(\ref{en_in})  we can conclude that
\begin{equation}
	   c_{1}2S^{\text{von}}(\left | \psi  \right \rangle )-1- \frac{1}{2^{\frac{n}{2}}}  \le S(\rho_{0,R} ^{u}) \le c_{1}2S^{\text{von}}(\left | \psi  \right \rangle )+1+\frac{1}{2^{\frac{n}{2}}}.
\end{equation}
Here, we use facts that operator EE of a pure state is twice as the von Neumann EE of the pure state and Shannon entropy of two variables is less than 1.

When the number of qubits is large, the operator EE can be well estimated as
\begin{equation}\label{oee_dev}
	\begin{split}
		S(\rho_{0,R} ^{u})  &\simeq c_{1}2S^{\text{von}}(\left | \psi  \right \rangle ) \\
		&= \frac{\alpha^{2}}{\alpha^{2} +\frac{1-\alpha^{2}}{2^{n}}}2S^{\text{von}}(\left | \psi  \right \rangle )\\ &=\frac{2}{1+\frac{1}{2^{n}\alpha^{2}}-\frac{1}{2^{n}}}S^{\text{von}}(\left | \psi  \right \rangle )\\
		&\simeq \frac{2}{1+\frac{1}{2^{n}\alpha^{2}}}S^{\text{von}}(\left | \psi  \right \rangle ).
	\end{split}
\end{equation}
Plugging $\alpha \simeq e^{-b_{0}pnt}$ into Eq.(\ref{oee_dev}) and averaging over RQCs, we can obtain
\begin{equation}
	\overline{S(\rho_{0,R} ^{u})} \simeq  \frac{2}{1+e^{2b_{0}pn(t-\frac{\text{In2}}{2b_{0}p})}} \overline{S^{\text{von}}(\left | \psi  \right \rangle )}.
\end{equation}

For $L_{1} \times L_{2}$ systems in this work, the typical growth of von Neumann entropy of $\left | \psi  \right \rangle $ during quantum chaotic dynamics without noise is \cite{PhysRevLett.112.011601,PhysRevX.9.021033,PhysRevX.7.031016}
\begin{equation}
	\overline{S^{\text{von}}(t)} \simeq  b_{1}L_{1}\cdot \min (b_{2}t, L_{2}),
\end{equation}
where $b_{1}$ and $b_{2}$ are constants. Here we have assumed $L_{1} \le L_{2}$ and choose cut at the middle of the systems. Therefore, we can write the operator EE of noisy RQCs as
\begin{equation}\label{mpoeeeq_AP}
	\overline{S(t,L_{1},L_{2},p)} \simeq  \frac{1}{1+e^{2b_{0}pn(t-\frac{\text{In2}}{2b_{0}p} )}}2b_{1}L_{1}\cdot \min (b_{2}t, L_{2}) ,
\end{equation}
As we can see in Eq.(\ref{mpoeeeq_AP}), the term $1/(1+e^{2b_{0}pn(t-\frac{\text{In2}}{2b_{0}p} )})$ is a Fermi–Dirac distribution-like function,
which is approximately a step function $\Theta (t)$ when $n$ is large,
\begin{equation}
	\Theta (t) = \left\{\begin{matrix}
		1 , t \leq \text{ln}2/(2b_{0}p)\\
		0 , t> \text{ln}2/(2b_{0}p).
	\end{matrix}\right .
\end{equation}
Therefore, the peak circuit depth and the corresponding operator EE of  Eq.(\ref{mpoeeeq_AP})
depend on whether $L_{2}/b_{2} \leq \text{ln}2/(2b_{0}p)$. When $L_{2}/b_{2} \leq \text{ln}2/(2b_{0}p)$, the peak circuit depth $t_{\text{p}}$ is $L_{2}/b_{2}$ and the corresponding operator EE is $2b_{1}L_{1}L_{2}$, whereas when $L_{2}/b_{2} > \text{ln}2/(2b_{0}p)$, the peak circuit depth is $\text{ln}2/(2b_{0}p)$ and the corresponding operator EE is $2b_{1}L_{1} \cdot b_{2}\text{ln}2/(2b_{0}p)$, i.e. the peak circuit depth is
\begin{equation}
	t_{\text{p}} \simeq \left\{\begin{matrix}
		L_{2}/b_{2},&L_{2} \leq \text{ln}2\cdot b_{2}/(2b_{0}p)\\
		\text{ln}2/(2b_{0}p),&L_{2} > \text{ln}2\cdot b_{2}/(2b_{0}p).
	\end{matrix}\right.
\end{equation}
and the corresponding maximum achievable operator EE is,
\begin{equation}
	S_{\text{max}} \simeq \left\{\begin{matrix}
		2b_{1}L_{1}L_{2},&L_{2} \leq \text{ln}2\cdot b_{2}/(2b_{0}p)\\
		\text{ln}2\cdot b_{1}b_{2} L_{1}/(b_{0}p),&L_{2} > \text{ln}2\cdot b_{2}/(2b_{0}p).
	\end{matrix}\right.
\end{equation}

\nocite{*}
\bibliography{ref.bib}

\end{document}